\documentclass[%
 aip,
 amsmath,amssymb,
 reprint,%
]{revtex4-1}

\usepackage{xcolor}
\usepackage{soul}
\usepackage{graphicx}
\usepackage{dcolumn}
\usepackage{bm}

\usepackage[utf8]{inputenc}
\usepackage[T1]{fontenc}
\usepackage{mathptmx}
\usepackage{etoolbox}
\makeatletter
\def\@email#1#2{%
 \endgroup
 \patchcmd{\titleblock@produce}
  {\frontmatter@RRAPformat}
  {\frontmatter@RRAPformat{\produce@RRAP{*#1\href{mailto:#2}{#2}}}\frontmatter@RRAPformat}
  {}{}
}%
\makeatother
\begin{document}

\preprint{AIP/123-QED}

\title[]{Inference of non-exponential kinetics through stochastic resetting}

\author{Ofir Blumer}
\affiliation{School of Chemistry, Tel Aviv University, Tel Aviv 6997801, Israel.}
\author{Shlomi Reuveni}
\affiliation{School of Chemistry, Tel Aviv University, Tel Aviv 6997801, Israel.}
\affiliation{The Center for Computational Molecular and Materials
Science, Tel Aviv University, Tel Aviv 6997801, Israel.}
\affiliation{The Center for Physics and Chemistry of Living Systems, Tel Aviv University, Tel Aviv 6997801, Israel.}
\author{Barak Hirshberg}
\email{hirshb@tauex.tau.ac.il}
\affiliation{School of Chemistry, Tel Aviv University, Tel Aviv 6997801, Israel.}
\affiliation{The Center for Computational Molecular and Materials
Science, Tel Aviv University, Tel Aviv 6997801, Israel.}
\affiliation{The Center for Physics and Chemistry of Living Systems, Tel Aviv University, Tel Aviv 6997801, Israel.}

\date{\today}

\begin{abstract}
We present an inference scheme of long timescale, non-exponential kinetics from Molecular Dynamics simulations accelerated by stochastic resetting. Standard simulations provide valuable insight into chemical processes but are limited to timescales shorter than $\sim 1 \mu s$. Slower processes require the use of enhanced sampling methods to expedite them, and inference schemes to obtain the unbiased kinetics. However, most kinetics inference schemes assume an underlying exponential first-passage time distribution and are inappropriate for other distributions, e.g., with a power-law decay. We propose an inference scheme that is designed for such cases, based on simulations enhanced by stochastic resetting. We show that resetting promotes enhanced sampling of the first-passage time distribution at short timescales, but often also provides sufficient information to estimate the long-time asymptotics, which allows the kinetics inference. We apply our method to a model system and a short peptide in an explicit solvent, successfully estimating the unbiased mean first-passage time while accelerating the sampling by more than an order of magnitude.
\end{abstract}

\maketitle

\section{Introduction}

Transitions between several metastable states are a key feature of many chemical and physical phenomena, such as chemical reactions and protein conformational changes. Molecular Dynamics (MD) simulations are an appealing computational tool for estimating the kinetic rates associated with such transitions. They track the dynamics of the system in microscopic detail, providing accurate predictions of both thermodynamic and kinetic properties. However, MD simulations require an integration timestep on the order of $\sim 1 \, \text{fs}$, limiting the simulations to a timescale of $\sim 1 \,\text{\textmu s}$. Processes that occur on longer timescales, such as protein folding and crystal nucleation, cannot be sampled efficiently in standard MD simulations~\cite{valsson_enhancing_2016,tiwary_metadynamics_2013,salvalaglio_assessing_2014,palacio-rodriguez_transition_2022,blumer_stochastic_2022}.

Different enhanced sampling methods were developed to overcome this well-known timescale problem. An incomplete representative list includes umbrella sampling~\cite{torrie_nonphysical_1977,kastner_umbrella_2011}, milestoning~\cite{faradjian_computing_2004,elber_milestoning_2020,elber2021modeling}, replica-exchange MD~\cite{SUGITA1999141,HANSMANN1997140,doi:10.1021/acs.jctc.7b00372}, Gaussian accelerated MD (GaMD)~\cite{doi:10.1021/acs.jctc.5b00436,wang2021gaussian,doi:10.1021/acs.jctc.0c00395}, and Metadynamics~\cite{barducci_metadynamics_2011,sutto_new_2012,valsson_enhancing_2016}. These methods promote extensive sampling of the transitions between metastable states within the accessible timescales of MD simulations, allowing the calculation of both thermodynamic averages and dynamic properties. For kinetics inference, most schemes assume that the underlying process has an exponential first-passage time (FPT) distribution~\cite{PhysRevLett.78.3908,doi:10.1021/acs.jctc.7b00372,doi:10.1021/acs.jctc.0c00395,tiwary_metadynamics_2013,salvalaglio_assessing_2014,palacio-rodriguez_transition_2022,khan2020fluxional,ray_kinetics_2023,valsson_enhancing_2016,blumer2024shorttime}, relying on transition rate theory (TST)~\cite{wigner1938transition,PETERS2017227} or Kramers' theory~\cite{KRAMERS1940284,PETERS2017435,EATR_2024,palacio-rodriguez_transition_2022}.

Exponential FPT distributions arise when a high energy barrier separates two narrow metastable states. 
However, many processes in nature have non-exponential FPT distributions. 
For instance, protein dynamics are sometimes better described by a sum of exponents with different rates, or skewed exponents~\cite{GRUEBELE2005701,ma2005kinetics,liu2008experimental,MOULICK2019807}. Power-law kinetics were also found experimentally for transitions between two stable conformers of an enzyme~\cite{grossman2018slow}. Even when the FPT distribution has an exponential tail, it can behave differently at short to intermediate times, for instance, obeying a power-law~\cite{doi:10.1021/acs.jpcb.7b07075}. Yet, kinetic inference schemes for processes with non-exponential FPT distributions are currently lacking. We propose such a scheme, designed for simulations enhanced by stochastic resetting (SR).

Resetting is the procedure of stopping random processes and restarting them subject to the sampling of independent, and identically-distributed initial conditions. It was shown to expedite different kinds of processes, including randomized computer algorithms~\cite{LUBY1993173,gomes_boosting_1998,montanari_optimizing_2002}, queuing systems~\cite{bressloff_queueing_2020,bonomo_mitigating_2022}, and various first-passage and search processes~\cite{evans_diffusion_2011,kusmierz_optimal_2015,Bhat_stochastic_2016,chechkin_random_2018,ray_peclet_2019,robin_random_2019,evans_run_2018,pal_search_2020,luo_anomalous_2022}. We recently demonstrated the power of resetting for enhanced sampling of MD simulations~\cite{blumer_stochastic_2022,blumer2024combining}. We showed that it can expedite rare events in molecular simulations when used as a standalone tool~\cite{blumer_stochastic_2022}, and in combination with Metadynamics simulations, which also improved the kinetics inference.~\cite{blumer2024combining}

The first two moments of the FPT distribution provide a sufficient condition for SR to be beneficial: if the ratio between the standard deviation and the mean, known as the coefficient of variation (COV), is greater than unity, introducing a small resetting rate is guaranteed to lower the mean FPT (MFPT)~\cite{Pal2022}. The COV of the exponential distribution is exactly equal to one, so processes with a broader FPT distribution can be accelerated with SR. However, they cannot be properly treated by most kinetics inference schemes from accelerated simulations, since these assume an exponential FPT distribution~\cite{tiwary_metadynamics_2013, salvalaglio_assessing_2014,palacio-rodriguez_transition_2022,khan2020fluxional,ray_kinetics_2023,blumer2024shorttime}. Resetting, on the other hand, can potentially provide accurate predictions, as it minimally perturbs the natural dynamics of the system between restart events. 

In this paper, we present a method to extract the unbiased MFPT of processes with non-exponential FPT distributions, from simulations accelerated by SR. We first present the theory underlying our method and use analytical FPT distributions to discuss its advantages and disadvantages. Next, we employ it to study a three-state, two-dimensional model system, and the dynamics of a nine-residues alanine peptide in explicit water. We show that our method can consistently predict the MFPT with high accuracy, with speedups of more than an order of magnitude over brute-force unbiased simulations.

\section{Theory}
\label{sec:theory}

Different protocols of resetting were suggested in the literature~\cite{evans2020stochastic}. Here we employ sharp resetting, where the waiting times between resetting events are constant, with some duration $T$, which is often called the timer. Sharp resetting was shown to provide greater acceleration than any other resetting protocol when performed with the optimal timer~\cite{Pal2017}. The MFPT $\langle \tau \rangle_T$ of a process with sharp resetting, as a function of the timer, is related to the unbiased FPT distribution through~\cite{eliazar2020mean}
\begin{equation}
\langle \tau \rangle_T = \frac{1}{1-S(T)} \int \limits_0^T S(t) dt,
\label{eq:sharpMFPT}
\end{equation}
with $S(t)$ being the survival probability of the FPT $\tau$ at time $t$,
\begin{equation}
S(t) = Pr(\tau > t).
\end{equation}
Note that Equation \ref{eq:sharpMFPT} requires the survival function only at $t \le T$. As the dynamics between resetting events remain unperturbed, simulations with some timer $T^*$ sample the exact survival probability at times $t \le T^*$. Given a sample of trajectories undergoing resetting with timer $T^*$, Equation \ref{eq:sharpMFPT} provides a practical tool to assess the MFPT with any timer $T<T^*$, indicating whether it is possible to further enhance the sampling by choosing shorter timers.

Simulations with timer $T^*$ also sample the exact conditional average $\langle \tau | \tau \le T^* \rangle$, i.e., the MFPT of trajectories with a FPT $\tau \le T^*$. Using the total expectation theorem, the unbiased MFPT can be expressed as
\begin{equation}
\langle \tau \rangle = \left(1-S(T^*)\right) \langle \tau | \tau \le T^* \rangle + S(T^*) \langle \tau | \tau > T^* \rangle.
\label{eq:trueMFPT}
\end{equation}
We note that simulations with timer $T^*$ provide all terms on the right-hand side of Equation
\ref{eq:trueMFPT}, except for $\langle \tau | \tau > T^* \rangle$, the MFPT of trajectories with a FPT $\tau > T^*$. If we could accurately estimate $\langle \tau | \tau > T^* \rangle$ through the behavior at times $t \le T^*$, we would be able to extract the unbiased MFPT via Equation \ref{eq:trueMFPT}. Fortunately, many FPT distributions have some distinctive decaying tail at $t>t'$, with $t'$ being some characteristic timescale. If we know that is the case, and choose a timer $T^*>t'$, we can sample part of this tail, fit it with the correct functional form, and obtain an accurate estimate of $\langle \tau | \tau > T^* \rangle$. 

For instance, if the FPT distribution decays exponentially with a rate $k$ at $t>t'$, then $\log \left( S(t) \right)$ would be linear for $t>t'$, with a slope $-k$. A linear fit of $\log \left( S(t)\right)$ in the proximity of $t=T^*>t'$ would provide $k$, and consequently 
\begin{equation}
\langle \tau | \tau > T^* \rangle = T^* + k^{-1}.
\label{eq:trueMFPTexp}
\end{equation}
Similarly, if the FPT distribution is governed by a power-law at $t>t'$, i.e $S(t) \propto t^{-\alpha}$ for $t>t'$ with some $\alpha>1$, then we can estimate~\cite{paretoDistribution}
\begin{equation}
\langle \tau | \tau > T^* \rangle = \frac{\alpha T^*}{\alpha-1}.
\label{eq:paretoPrediction}
\end{equation}
Once we estimate $\langle \tau | \tau > T^* \rangle$, we substitute it in Equation \ref{eq:trueMFPT} and have an estimate of the unbiased MFPT.
In practice, when using real data, for both exponential and power-law tails, we fit the log of the survival function for every choice of $t' < T^*$ and choose the $t'$ that provides the best linear fit, i.e., with the highest Pearson correlation coefficient. Note that for an exponential and power law tail, the linear fit should be done on semi-log and log-log scales, respectively. Ideally, $T^*$ should be greater than $t'$, but not too large so that sampling by resetting would still lead to acceleration over standard MD simulations. In all the examples below, we find that it is possible to estimate the scaling of the distribution tails through this procedure while benefiting from significant speedups.

\begin{figure*}
  \centering
  \includegraphics[width=0.8\linewidth]{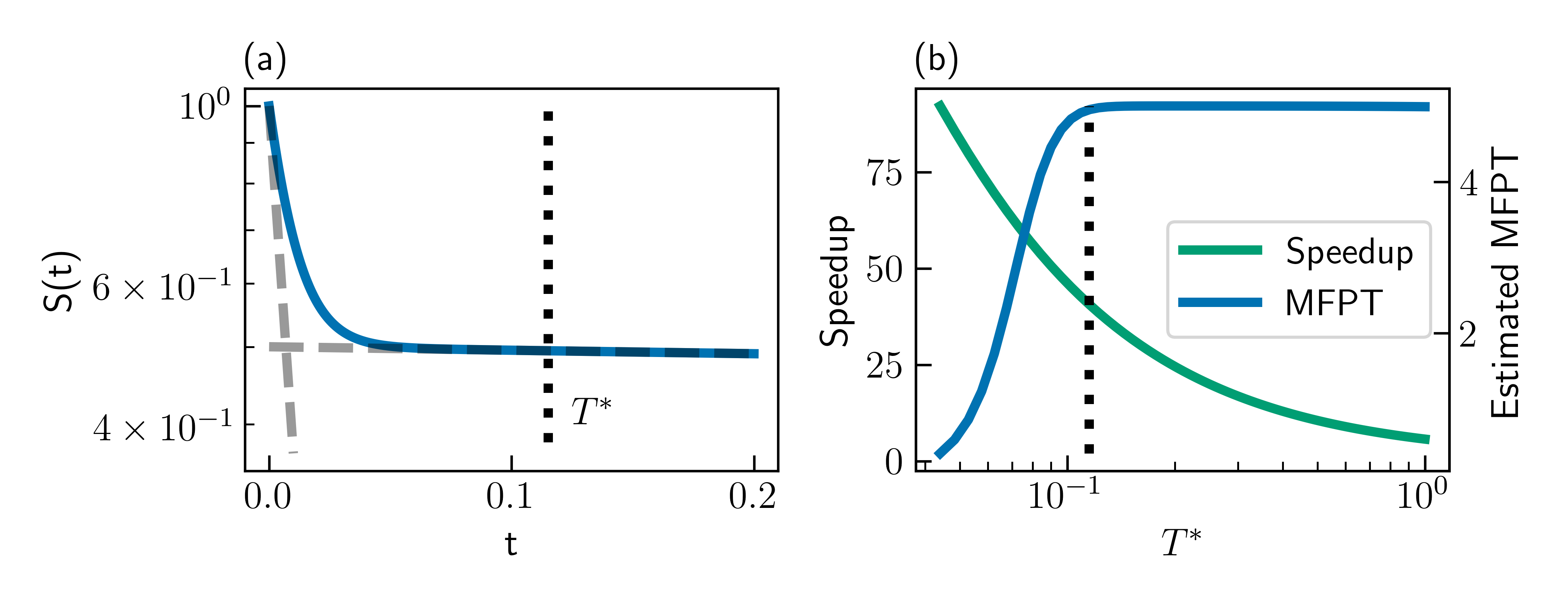}
  \caption{
    (a) Exact survival function for a hyperexponential distribution (Equation \ref{eq:twoExp}). 
  The dashed gray lines show slopes of $-k_1$, $-k_2$ for comparison. The dotted black line highlights a specific $T^*=0.115$. (b) Speedup (green) and estimated MFPT (blue) as a function of the timer using Equations \ref{eq:trueMFPT},\ref{eq:trueMFPTexp} and the analytical survival function. The dotted black line highlights a timer of $T^*=0.115$.}
  \label{fig:twoExponentsSurvivalSpeedup}
\end{figure*}

\section{Results and discussion}
\label{sec:results}
\subsection{Anlytical FPT distributions}
\paragraph{Hyperexponential distribution.}
\label{sec:illu}

We first employ our approach to an analytical, dimensionless FPT distribution, a hyperexponential distribution composed of two exponents with rates $k_1$ and $k_2$,
\begin{equation}
p(\tau) = A k_1 e^{-k_1\tau}+ (1-A)k_2 e^{-k_2\tau}.
\label{eq:twoExp}
\end{equation}
We take $A = 0.5$ for simplicity and $k_1 = 100 \gg k_2 = 0.1$ for a separation of timescales between the terms.
The survival function is plotted in Figure \ref{fig:twoExponentsSurvivalSpeedup}(a). It shows fast decay at short times and fits the slower rate $k_2$ at longer times. The rate of decay is $\sim 1\%$ off $k_2$ at time $T^*$, marked with a black dashed line. The gray dotted lines decay with rates $k_1$ and $k_2$. 

For this analytical example, we know the exact survival function and can use Equation \ref{eq:sharpMFPT} to obtain the exact MFPT under resetting for any choice of timer $T^*$. 
We can also estimate the MFPT with no resetting ($=5.005$) through Equations \ref{eq:trueMFPT}-\ref{eq:trueMFPTexp}, using the exact values of $\langle \tau | \tau \le T^* \rangle$ and $S(T^*)$, and evaluating $k = -\frac{d \log(S(t))}{dt} \vert_{t=T^*}$ with the exact derivative. 
Figure \ref{fig:twoExponentsSurvivalSpeedup}(b) shows the resulting estimations as a function of $T^*$ (blue, right y-axis). It also presents the speedup (green, left y-axis), with the speedup defined as the ratio between the MFPT without and with resetting, respectively. The dashed black line marks the results with the specific timer $T^*$ shown in Figure \ref{fig:twoExponentsSurvivalSpeedup}(a). We observe that the MFPT estimation is within $1\%$ of the true value for speedups up to $\sim 40$. This is expected since the slope of the survival at $T^*$ is within $1\%$ of $k_2$ at $T^*$. This means that, for a process with this FPT distribution, we could spend 40 times less computational time per first-passage event, compared to tedious, unbiased simulations, and obtain nearly the same accuracy. 

However, in more realistic scenarios, the accuracy of the predicted MFPT would also be limited by the number of trajectories sampled and how well they capture the slope of the distribution tails. We investigate the sensitivity of our approach to sampling noise by numerical sampling from the distribution in Equation \ref{eq:twoExp} to estimate its survival function.
To estimate the MFPT in this case, we sampled 1000 batches for every timer, each composed of 100 FPTs that were numerically sampled from the distribution. 
We constructed trajectories with resetting in the following way: We first sampled a time from the FPT distribution without resetting and compared it with the timer. If the timer was smaller, we tallied it up and sampled a new time from the FPT distribution without resetting. We repeated the process until the sample was smaller than the timer. In that case, we added the sample to the sum, and the sampling of the trajectory was completed. The total summed time was a sample of the FPT with resetting at that value of $T^*$.

\begin{figure}
  \centering
  \includegraphics[width=0.8\linewidth]{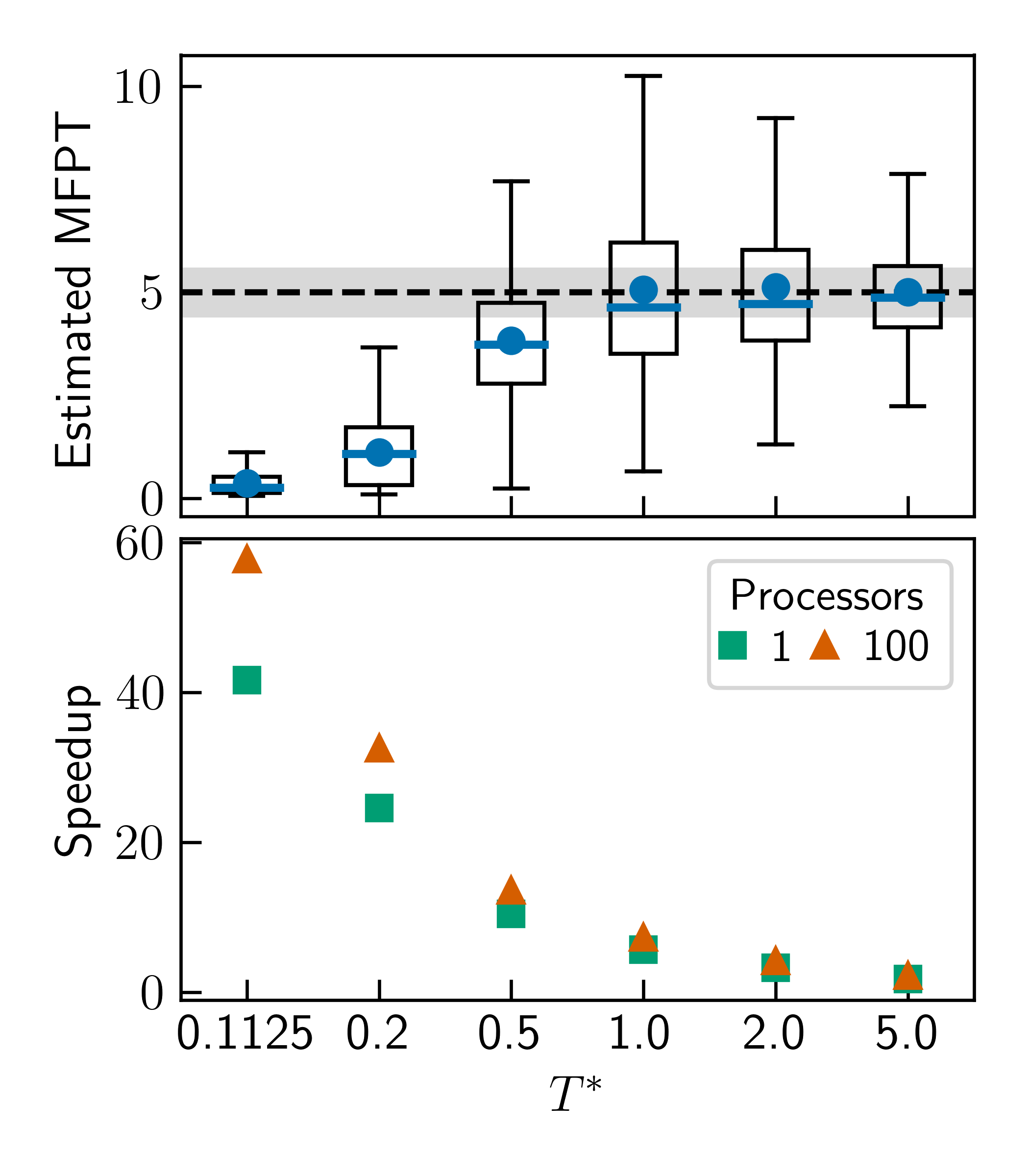}
  \caption{Estimated MFPT (top panel) and speedup (bottom panel) as a function of $T^*$, for the hyperexponential distribution. The circles, horizontal lines, boxes, and whiskers in the top panel show the mean, median, IQR, and extreme values within 1.5 IQR of the first and third quartiles, respectively. The shaded region shows the IQR of 1000 batches of 100 unbiased simulations each. The speedup was calculated for a single processor (green squares) and 100 processors (orange triangles).}
  \label{fig:twoExponentsSpeedup}
\end{figure}

Figure \ref{fig:twoExponentsSpeedup} shows the MFPT estimation as a function of the selected timer $T^*$ (top panel). The boxes show the range between the first and third quartiles (interquartile range, IQR), and the whiskers show extreme values within 1.5 IQR below and above these quartiles. The circles and horizontal lines give the mean and median, respectively. The associated speedups are plotted with green squares in the bottom panel. We find that a timer of $0.115$ gives a speedup of $40$, as anticipated, but provides a poor estimate of the unbiased MFPT, due to insufficient sampling in the proximity of $t'$. Longer timers provide better estimations, with accurate averages ($<0.6\%$ absolute deviation) for speedups of up to $\sim 6$. Increasing the number of samples improves the estimations: Supplementary Figure 1. shows results equivalent to those of Figure \ref{fig:twoExponentsSpeedup} but sampling 1000 first-passage events for each timer. We find that a timer of $0.2$ already gives $<4\%$ absolute deviation, with a speedup of $\sim 24$. 

\paragraph{Additional speedup by parallelization.}

So far, we assumed that all simulations are performed serially, each one initiated only when the former is done. This would be the case if only one processor was available. But we can usually perform simulations in parallel on several processors. This becomes increasingly affordable with the improvement in computer power, as demonstrated, for instance, by the parallel replica dynamics (ParRep) method~\cite{voter1998parallel,PEREZ201590,doi:10.1021/acs.jctc.5b00916}. 
As explained in Ref.~\cite{PEREZ201590}, the improvement in computer power is mainly reflected in ever-greater levels of parallelization, and not in per-processor speed. While it does not directly solve the timescale problem, it greatly benefits parallelizable enhanced sampling methods.

Consider, for instance, running 100 trajectories on 100 processors simultaneously: for unbiased statistics, one has to wait for all trajectories to end. Thus, the walltime of the sampling is the FPT of the longest trajectory, which is always larger than the empirical MFPT. For the hyperexponential distribution above, using 100 trajectories, it is $\sim45$, i.e., almost an order of magnitude greater than the MFPT (see Equation S1). The larger the COV, i.e., the larger the fluctuations in the FPT distribution, the larger the walltime compared to the MFPT. Resetting lowers the COV~\cite{reuveni2016optimal}, reducing the ratio between the longest first-passage time and the MFPT. This, in turn, translates to greater speedups in parallelizable settings. We demonstrate this by plotting the walltime speedup with 100 processors for the hyperexponential distribution (orange triangles in Figure \ref{fig:twoExponentsSpeedup}). The walltime speedup is defined as the ratio between the sampling walltimes without and with resetting, respectively. We find that in this case, it is larger by up to $40 \%$ than the equivalent speedup with a single processor. 

\paragraph{Power law distribution.}

We finish this section with a non-exponential analytical example, whose behavior is qualitatively different than the previous example: The Pareto distribution,
\begin{equation}
p(\tau) = \begin{cases} \frac{\sigma \tau_m^\sigma}{\tau^{\sigma+1}} & \tau \ge \tau_m \\ 0 & \tau < \tau_m, \end{cases}
\end{equation}
with $\sigma = 1.25$ and $\tau_m = 1$. It has power-law decay (Equation \ref{eq:paretoPrediction} is thus used for MFPT estimations), an MFPT of $5$, and its COV without resetting diverges. For this case, we find that the walltime speedup using 100 processors is $\sim 15$, an order of magnitude larger than that using a single processor ($\sim 2$).
Since the power law decay starts at $\tau_m$, all FPT measurements with timers greater than $\tau_m$ sample the tail. Moreover, since the decay is uniform, different timers give very similar MFPT estimations. We thus present results for a single timer $T^*=2$. 

\begin{figure}[t!]
  \centering
  \includegraphics[width=0.8\linewidth]{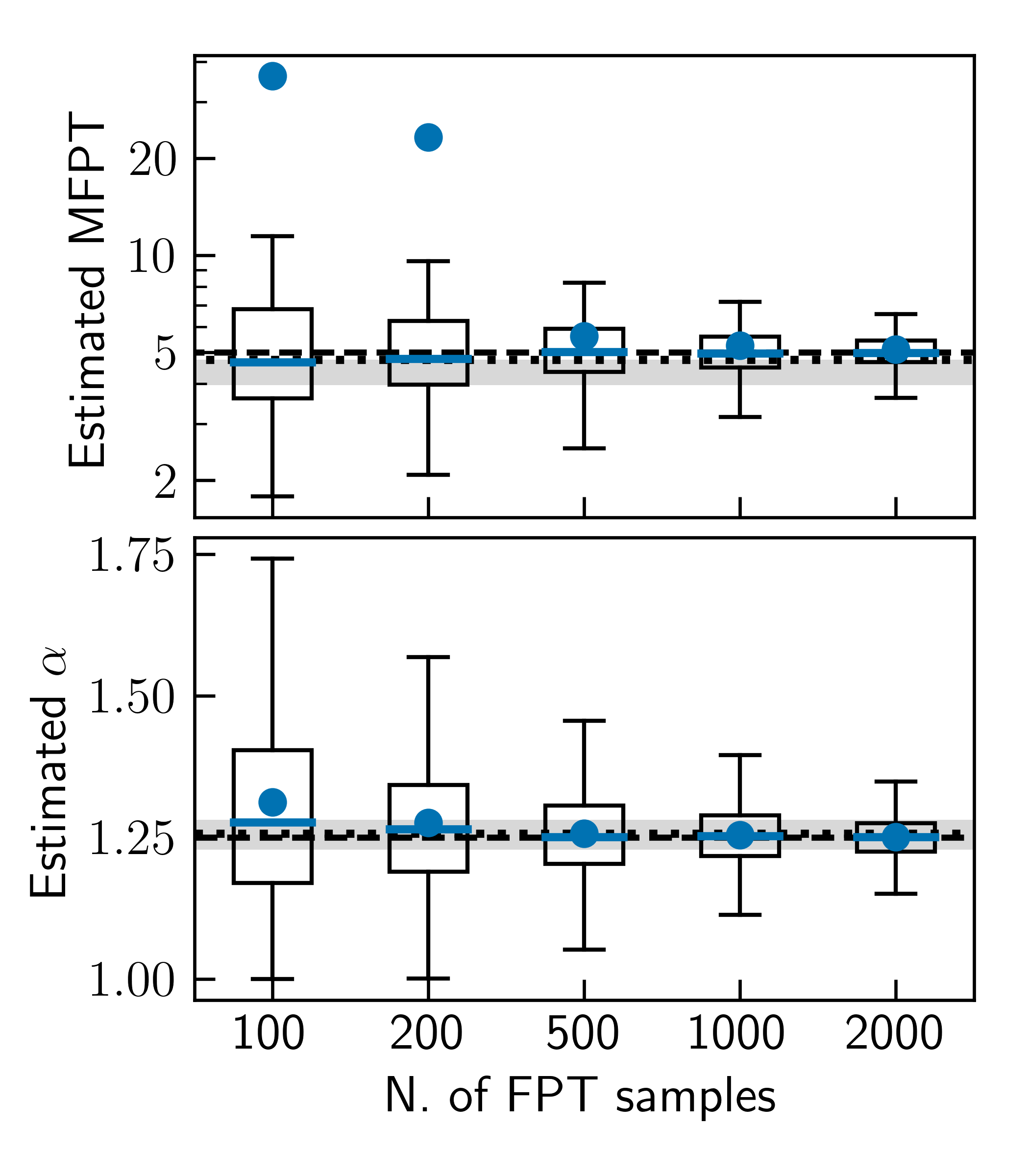}
  \caption{
 The Pareto distribution: Estimated MFPT (top) and estimated $\alpha$ (bottom), as a function of the number of first-passage samples. The dashed black lines give the analytic MFPT and $\alpha$ in the corresponding panels.
 The circles, horizontal lines, boxes, and whiskers show the mean, median, IQR, and extreme values within 1.5 IQR of the first and third quartiles, respectively. The dotted lines and the shaded regions show the mean and IQR, respectively, for 1000 sets of 2000 unbiased measurements each.}
  \label{fig:pareto}
\end{figure}

Figure \ref{fig:pareto} shows the estimations of the MFPT with $100-2000$ FPT samples (top panel). The median of the MFPT estimations is close to the true MFPT even with 100 samples with resetting. The mean is overestimated due to a few outlier values not shown. With $500$ and $2000$ samples, the mean is only $\sim 12 \%$ and $\sim 2 \%$ off the true MFPT, respectively. 
The bottom panel shows the estimates of $\alpha$ from the same data used in the top panel, demonstrating that we can accurately obtain power-law tails from simulations with resetting.

\subsection{Three-state model}

\begin{figure*}[t!]
  \centering
  \includegraphics[width=0.8\linewidth]{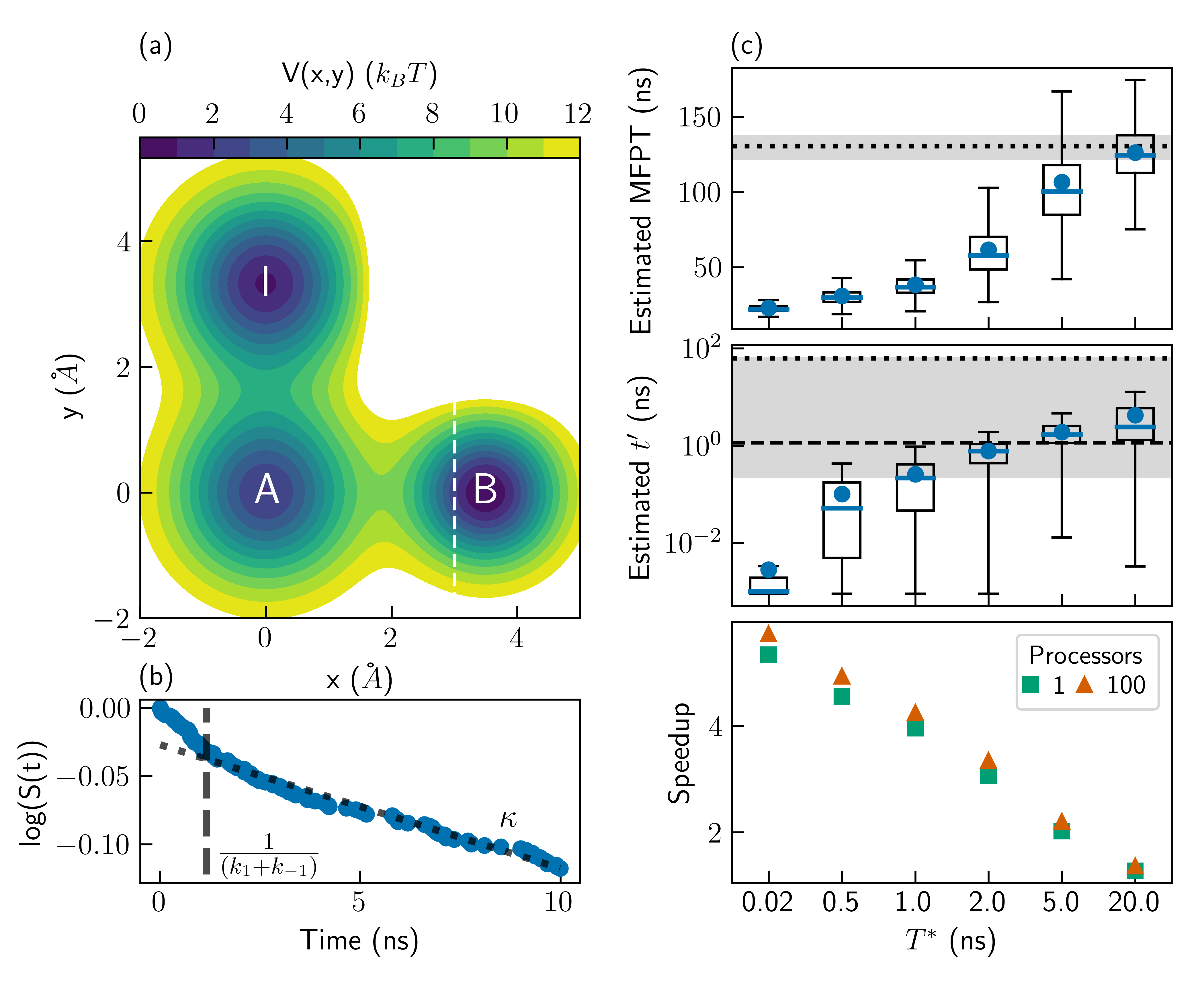}
  \caption{(a) The three-state system. The white dashed line marks the first-passage criterion. (b) The survival function at times $ < 10 \, ns$. The dashed line marks the time $t=\frac{1}{\left(k_1+k_{-1}\right)}$ and the dotted line shows decay at rate $\kappa=\frac{k_{-1}}{k_1+k_{-1}} k_2$. (c) Estimated MFPT (top), timescale $t'$ (center), and speedup (bottom) as a function of the timer. Speedup was calculated for simulations on a single processor (green squares) or on 100 parallel processors (orange triangles). The circles, horizontal lines, boxes, and whiskers show the mean, median, IQR, and extreme values within 1.5 IQR of the first and third quartiles, respectively. The dotted lines and shaded areas show mean values and IQR for 100 bootstrapping sets of 100 unbiased simulations each, respectively (1000 independent unbiased trajectories were collected in total). The dashed line in the middle panel shows $t=\frac{1}{\left(k_1+k_{-1}\right)}$.}
  \label{fig:threeStates}
\end{figure*}

We next apply our inference scheme for MD simulations of a particle diffusing over a two-dimensional three-state potential (Figure \ref{fig:threeStates}(a), Equation S2). This potential was introduced by Khan et al. to represent a simple kinetic model, with two first-order reactions~\cite{khan2020fluxional}. The particle is initially positioned at state A. One reaction path leads to state I, with a kinetic rate $k_1$. It is a reversible reaction, with transitions from I to A governed by rate $k_{-1}<k_1$. The second path leads to the product state B with rate $k_2<k_{-1}$. As Khan et al. point out, the system should reach a quasi-equilibrium between states A and I at time $\left(k_1+k_{-1}\right)^{-1}$, after which, transitions from states A and I to B follow an effective rate $\kappa=\frac{k_{-1}}{k_1+k_{-1}} k_2$. We assessed the values of $k_1$, $k_{-1}$, and $k_2$ using unbiased simulations in which only a single reaction path was available, with the other blocked by a strong repulsive potential. The survival function shown in Figure \ref{fig:threeStates}(b) confirms that transitions to state B follow rate $\kappa$ at times larger than some $t'$, while faster at $t < t'$. 

Figure \ref{fig:threeStates}(c) presents estimations of the MFPT (top panel) and time $t'$ (center panel) as a function of $T^*$, sampling 100 first-passage events per batch for each timer. The associated speedups are plotted in the bottom panel. The speedup is similar with 1 and 100 processors because the COV without resetting is $\sim 1.05$, very close to unity. We find that even the shortest timer, providing speedups of $\sim 6$, estimates the MFPT within an order of magnitude of the unbiased value. As expected, the accuracy improves with larger timers. In addition, unbiased simulations fail to provide any insight on the timescale $t'$: the IQR is very broad, and the mean is higher than the third quartile due to large outlier values. But, simulations with resetting correctly estimate its order of magnitude with almost any timer choice.

\subsection{Alanine peptide}

\begin{figure*}[t!]
  \centering
  \includegraphics[width=0.8\linewidth]{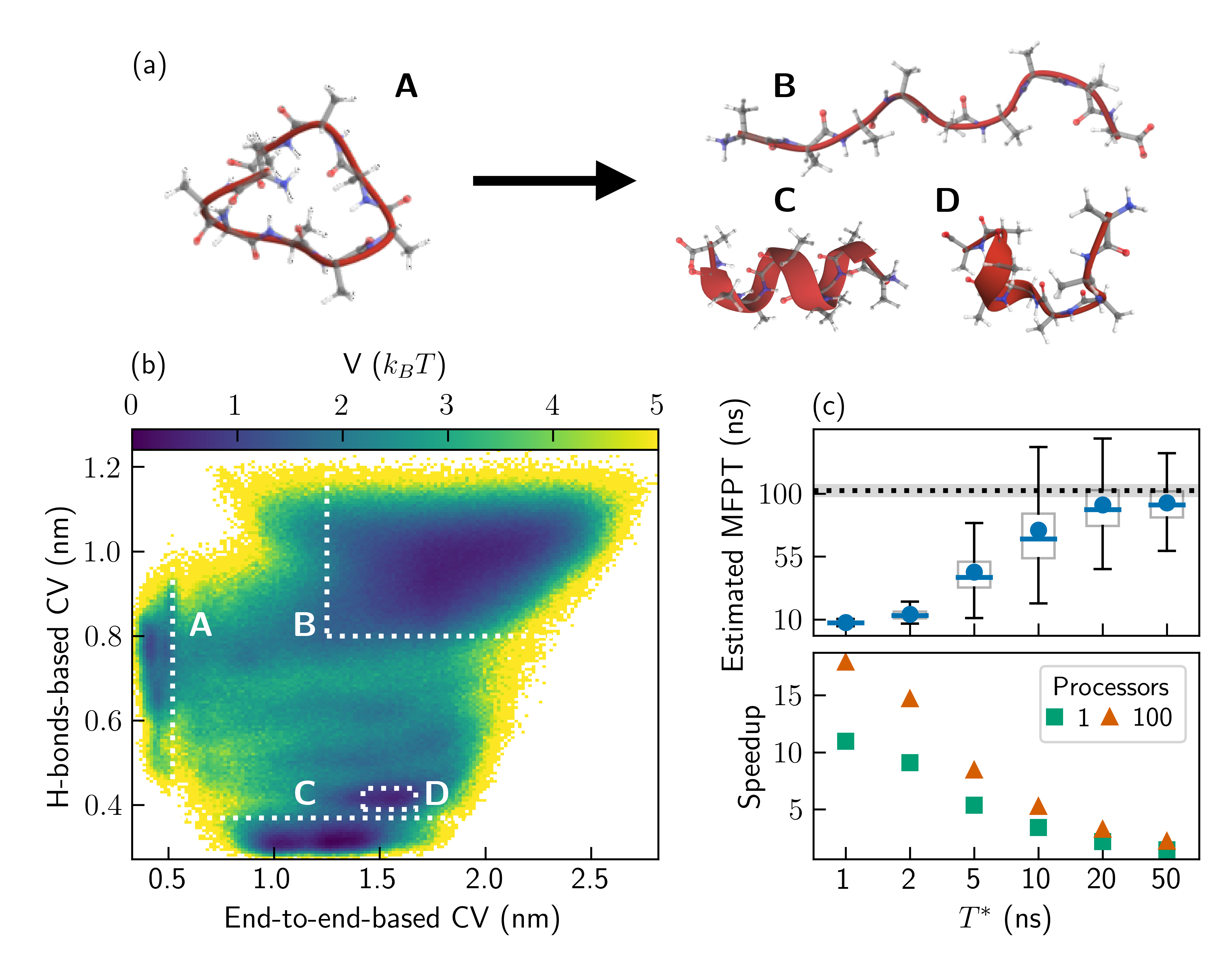}
  \caption{(a) Ball-and-stick and cartoon representations of four stable configurations of a nine-residues alanine peptide. The white, gray, blue, and red spheres represent hydrogen, carbon, nitrogen, and oxygen atoms, respectively. (b) Free energy along an end-to-end-based CV and an H-bonds-based CV. The white dotted lines define four metastable states. (c) Estimated MFPT (top) for different timers. The MFPT of 1000 unbiased trajectories is indicated with a black dotted line in the top panel, with the gray shading showing $\pm 1/\sqrt{1000}$ the standard deviation -- the estimated error. The circles, horizontal lines, boxes, and whiskers show the mean, median, IQR, and extreme values within 1.5 IQR of the first and third quartiles, respectively. The speedup using a single or 100 processors is plotted in the bottom panel with green squares and orange triangles, respectively.}
  \label{fig:ala9}
\end{figure*}

Many proteins are characterized by a rugged free-energy surface (FES), with multiple metastable states separated by low energy barriers~\cite{MOULICK2019807,nevo2005direct,wolynes1996fast}.
An extreme case is downward folding, where there are no energy barriers along the folding path~\cite{GRUEBELE2005701,ma2005kinetics,liu2008experimental}. 
The assumption of Poisson statistics, which is usually acceptable in systems with high energy barriers, is often invalid in protein dynamics. Therefore, we expect protein dynamics to be a natural proving ground for our new method. To demonstrate it, we employ our method to study the dynamics of a nine-residues peptide of alanine (Figure \ref{fig:ala9}(a)), which is the shortest peptide known to form a stable $\alpha$-helix~\cite{ayaz2021non}. It also forms a stable ``loop'', similar to other alanine chains~\cite{doi:10.1021/acs.jpcb.7b07075}. 
This system was chosen to benchmark our approach due to its multiple transitions between metastable states on a long timescale, but not too slow, allowing benchmarking against brute-force unbiased simulations.

Figure \ref{fig:ala9}(b) shows the FES of the system along two collective variables (CVs), obtained from a continuous, 2 \textmu s-long unbiased trajectory. One-dimensional FES along these CVs are provided in Supplementary Figure 2. The first CV is the end-to-end distance $x$, calculated using the center of mass of the two edge residues, which identifies the closed ``loop'' state (A, $x<0.52 \,\text{nm}$). The second is the average of three distances between pairs of H-donor nitrogen and acceptor oxygen within the peptide. This average, denoted here as $y$, identifies the helix~\cite{ayaz2021non} (C, $y<0.37\,\text{nm}$). We also identify a broad basin of metastable open configurations (B, $x>1.25 \, \text{nm and} \, y > 0.8 \, \text{nm}$) and a metastable intermediate state, where two of the three H-bonds are formed (D, $1.42<x<1.67 \, \text{nm and} \,  0.39<y<0.44 \, \text{nm}$). Representative configurations of the states are presented in Figure \ref{fig:ala9}(a).

We first sampled 100 configurations of state A, obtained in time intervals of $0.5 \, \text{ns}$ from a trajectory restricted to state A using a strongly repelling potential (see Methods section) along the end-to-end-based CV. We then ran independent simulations with resetting, uniformly sampling the initial conditions from those configurations. The first-passage was defined as settling into any other stable state: B, C, or D. For comparison, we also performed 1000 brute-force unbiased simulations to determine that this process has a MFPT of $\sim 102 \, \text{ns}$ and a COV of $\sim 1.46$. The tail of the FPT distribution is exponential, hence, we use Equation \ref{eq:trueMFPTexp} for the MFPT estimations.

Figure \ref{fig:ala9}(c) shows the MFPT estimations from simulations with resetting with different timers (top). The associated speedups are plotted in the bottom panel. The shortest timer provides speedups of $\sim 11$ and $\sim 18$ using 1 and 100 processors, respectively. Using this timer, we estimate the MFPT as $7.9 \, \text{ns}$, about an order of magnitude lower than the true value. Larger timers give more accurate results and still lead to accelerations. For instance, using a timer of 20 ns we estimate the MFPT as $\sim 92 \, \text{ns}$ with speedups of $\sim 2.2$ and $\sim 3.3$ using 1 and 100 processors, respectively. 

\section*{Conclusions}

To conclude, we presented an inference scheme for non-exponential kinetics from MD simulations accelerated by resetting. Almost all kinetics inference methods for enhanced sampling simulations assume an underlying exponential FPT distribution, but in many cases of interest, this assumption simply does not hold. Resetting is an especially appealing tool for non-exponential systems since it is guaranteed to expedite processes whose FPT distribution is broader than exponential. Moreover, it does not affect the dynamics between restarts, and samples the natural dynamics of the underlying process up to the resetting time. If the FPT distribution has a well-behaved asymptotic decay starting at times that are slightly shorter than the resetting time, we can estimate the tail of the distribution by its behavior at shorter timescales. This, in turn, allows estimating the unbiased MFPT of non-exponential processes using simulations accelerated by resetting.

Our approach becomes increasingly favorable as the number of available processors grows. With standard unbiased simulations, a few trajectories with long FPTs dominate the total walltime for all simulations to end, regardless of the number of available processors. By limiting the trajectories to short timescales using resetting, we use the available resources more efficiently. Compared to sets of unbiased trajectories, we obtain similar accuracy at much shorter simulation times. As acknowledged earlier by the developers of ParRep, parallelizability is an increasingly important quality: while the rapid improvement in computer power does not enable longer simulations, per se, it makes trivially parallelizable methods much more appealing~\cite{doi:10.1021/acs.jctc.5b00916,PEREZ201590}.

We applied our method to several analytical FPT distributions, a three-state model potential and an alanine peptide in explicit water. We obtain speedups of more than an order of magnitude with small statistical errors in the predicted MFPT. In both systems, we rely on the fact that the FPT distribution has either an exponential or power-law tail. However, our approach is much more general and can be employed to other kinds of distributions.

\section*{Methods}
\label{sec:methods}

Samples from analytical distributions were obtained using Python. The script is available on the associated GitHub repository.
Simulations of the three-states model system were carried out in the Large-scale Atomic/Molecular Massively Parallel Simulator (LAMMPS)~\cite{LAMMPS}, following the motion of a single particle with a mass of $40\, \text{gr}\cdot \text{mol}^{-1}$. They were performed in the canonical (NVT) ensemble at a temperature of $300 \, K$, using a Langevin thermostat~\cite{schneider_molecular-dynamics_1978} with a friction coefficient of $0.01 \, \text{fs}^{-1}$. 

 We used input files by Ayaz~\cite{Ayaz2021} for the alanine peptide. The simulations were carried out in GROMACS 2019.6~\cite{abraham_gromacs_2015}, using the Amber03 force field~\cite{duan2003point} for the peptide and the extended simple point-charge (SPC/E) model~\cite{doi:10.1021/j100308a038} for the solvent. They were performed in the NVT ensemble at $300 \, K$ using a stochastic velocity rescaling thermostat~\cite{bussi_canonical_2007}. Additional simulation settings are as reported in Ref.~\cite{ayaz2021non}.

The integration timestep was $1 \, \text{fs}$ for both systems. The progress of the systems in CV space was measured using PLUMED 2.7.1~\cite{bonomi_plumed_2009,tribello_plumed_2014,Bonomi2019}. For the alanine peptide, we tested whether the system entered a new basin every $1 \, \text{ps}$. A first-passage event was considered only if the system was observed in the new basin for at least 5 consecutive tests. We also used PLUMED to restrict the system to state A when obtaining initial configurations. We added a harmonic potential $0.5k\left(x-0.52\right)^2$ at $x>0.52\,\text{nm}$, with $k=100\, k_BT/\text{nm}^2$ and $x$ being the end-to-end-based CV.

Lastly, in Figures 2-4, we report statistics over 1000 independent sample batches. In Figures 5-6, we present statistics over 1000 bootstrapping sets. The total number of simulations used for bootstrapping, and the total number of trajectories ending in first-passage for each timer, are given in Supplementary Tables 1-2.

\begin{acknowledgments}
B. H. acknowledges support from the Israel Science Foundation (grants No. 1037/22 and 1312/22) and the Pazy Foundation of the IAEC-UPBC (grant No. 415-2023). This project has received funding from the European Research Council (ERC) under the European Union’s Horizon 2020 research and innovation program (grant agreement No. 947731).
\end{acknowledgments}

\section*{Data Availability}

Example input files, source data for all plots, and a Python class for the proposed inference method are available in the GitHub repository: 

\noindent \href{https://github.com/OfirBlumer/nonExponentialKinetics}{\url{github.com/OfirBlumer/nonExponentialKinetics}}.

\bibliography{bibli}

\end{document}


\section{Results for the hyperexponential distribution with 1000 samples per timer}

\begin{figure}[H]
  \includegraphics[width=0.5\linewidth]{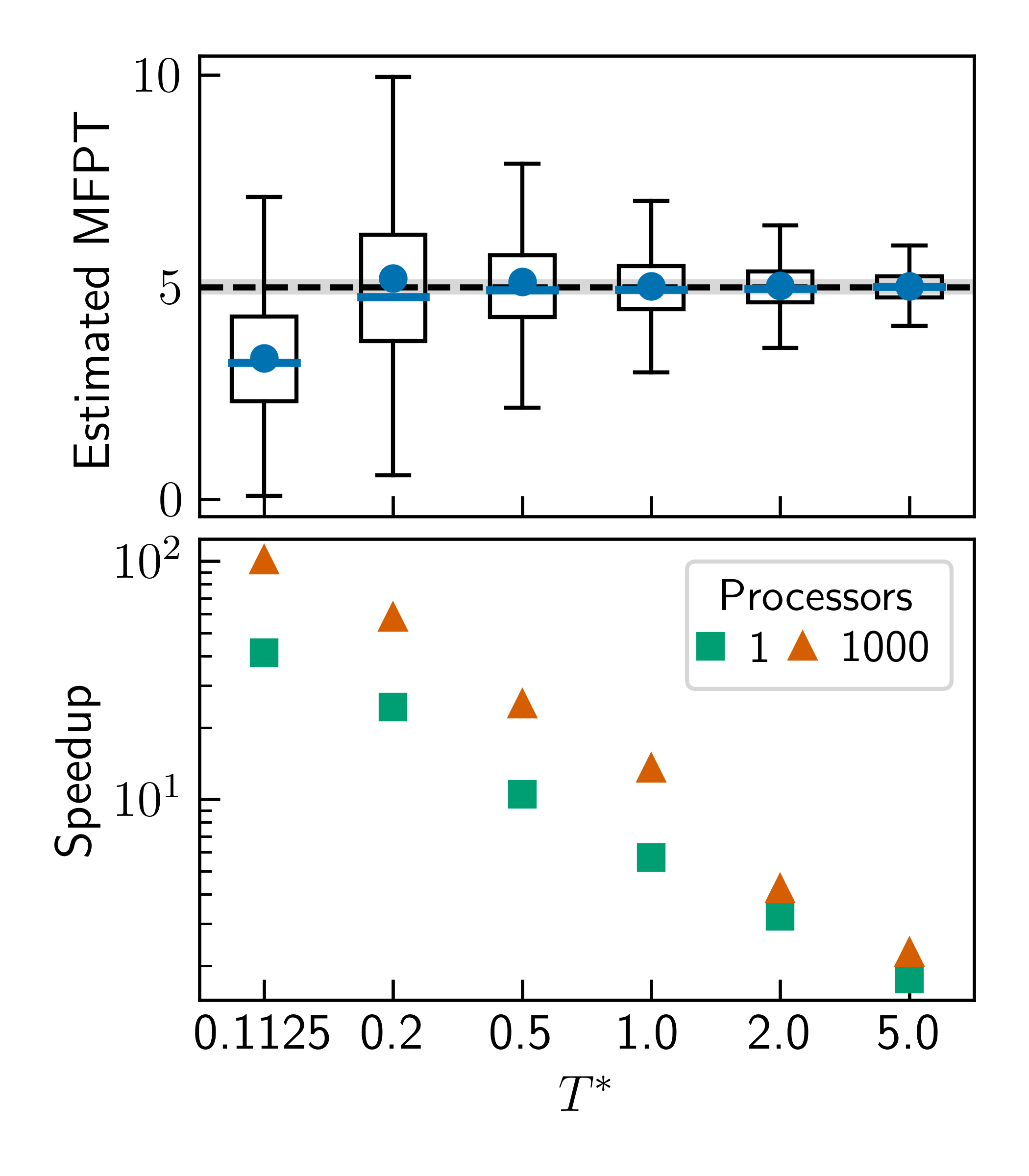}
  \caption{Estimated mean first-passage time (MFPT) (top panel) and speedup (bottom panel) as a function of $T^*$, for the hyperexponential distribution, with 1000 samples per batch per timer. The circles, horizontal lines, boxes, and whiskers in the top panel show the mean, median, interquartile range (IQR), and extreme values within 1.5 IQR of the first and third quartiles, respectively. The shaded region shows the IQR of 1000 batches of 1000 unbiased simulations each. The speedup was calculated for a single processor (green squares) and 1000 processors (orange triangles).}
\label{fig:twoexp1000}
\end{figure}

Here we provide a figure equivalent to Figure 2 of the manuscript but with 1000 first-passage samples for each timer instead of 100 (Figure \ref{fig:twoexp1000}). We find accurate results with a timer $T^* = 0.2$, providing speedups of $\sim 24$ and $\sim 58$ over simulations without resetting, using 1 and 1000 processors, respectively.

\section{Expected walltime with multiple processors}

The average walltime of sampling $N_p$ trajectories ending in first-passage by running $N_p$ simulations in parallel on $N_p$ processors is 
\begin{equation}
\left\langle \tau_{wall}(N_p) \right\rangle = N_p \int \limits_{0}^\infty \tau f(\tau) 
F(\tau)^{N_p-1}d\tau.
\label{eq:walltime}
\end{equation}
Here, $f(\tau)$ is the first-passage time (FPT) probability density, and $F(t) = \int \limits_0^t f(\tau)d\tau$ is the cumulative distribution function.

This equation was derived through the following arguments: The probability of a specific trajectory to end in first-passage on time $\tau$ is $f(\tau)$ by definition. It is the largest FPT only if all other $N_p -1$ trajectories ended on time $<\tau$. Since the probability of a single trajectory to have FPT $<\tau$ is $F(\tau)$, the probability of all $N_p -1$ independent trajectories to have FPT $<\tau$ must be $F(\tau)^{N_p-1}$. Thus, we find that the probability of a specific trajectory to end on time $\tau$ after all other trajectories ended is $f(\tau) F(\tau)^{N_p-1}$. Because we are not interested in the probability of a specific trajectory, and any of the $N_p$ trajectories may have the largest FPT, the probability of the largest FPT of all trajectories being $\tau$ is $N_p f(\tau) F(\tau)^{N_p-1}$. To obtain the expectation value of the largest FPT, i.e. the walltime, we have to integrate over all possible $\tau$ times their probability. This is exactly what Equation \ref{eq:walltime} does. 

\section{The three-state potential}

Here, we provide the details of the three-state potential. Its form is given in Equation \ref{eq:threeState}, with the following parameters: $a=11.5$, $b=13.5$, $c=12.665$, $s_a=1$, $s_b=0.8$ and $s_c=1$. They fit distances in units of $\AA$ and energy in units of $1 \, k_B T$.

\begin{equation}
    \label{eq:threeState}
\begin{split}
    V\left(x,y\right) = -a\exp\left(-\frac{x^2+y^2}{2s_a^2}\right)-b\exp\left(-\frac{\left(x-3.5\right)^2+y^2}{2s_b^2}\right)-\\
    c\exp\left(-\frac{x^2+\left(y-3.34\right)^2}{2s_c^2}\right)+
    \frac{2.68^{10}}{2\left[\left(x-4\right)^{10}+\left(y-4\right)^{10}\right]}
\end{split}
\end{equation}

As in Ref.~\cite{khan2020fluxional}, we added harmonic constraints at $x < -2\, \AA$, $x > 5\, \AA$, $y < -2\, \AA$ and $y > 5\, \AA$ to trap the diffusing particle within the Gaussian wells. The constraints have the form $0.5k\left(q-q_0\right)^2$, with $k=50$, $q$ being either $x$ or $y$, and $q_0 \in \{-2,5\}$.

\section{One-dimensional free energy surfaces of alanine peptide}

Here we provide one-dimensional free energy surfaces (FES) of the nine-residues alanine peptide along the two chosen collective variables (CVs).

\begin{figure}
  \includegraphics[width=\linewidth]{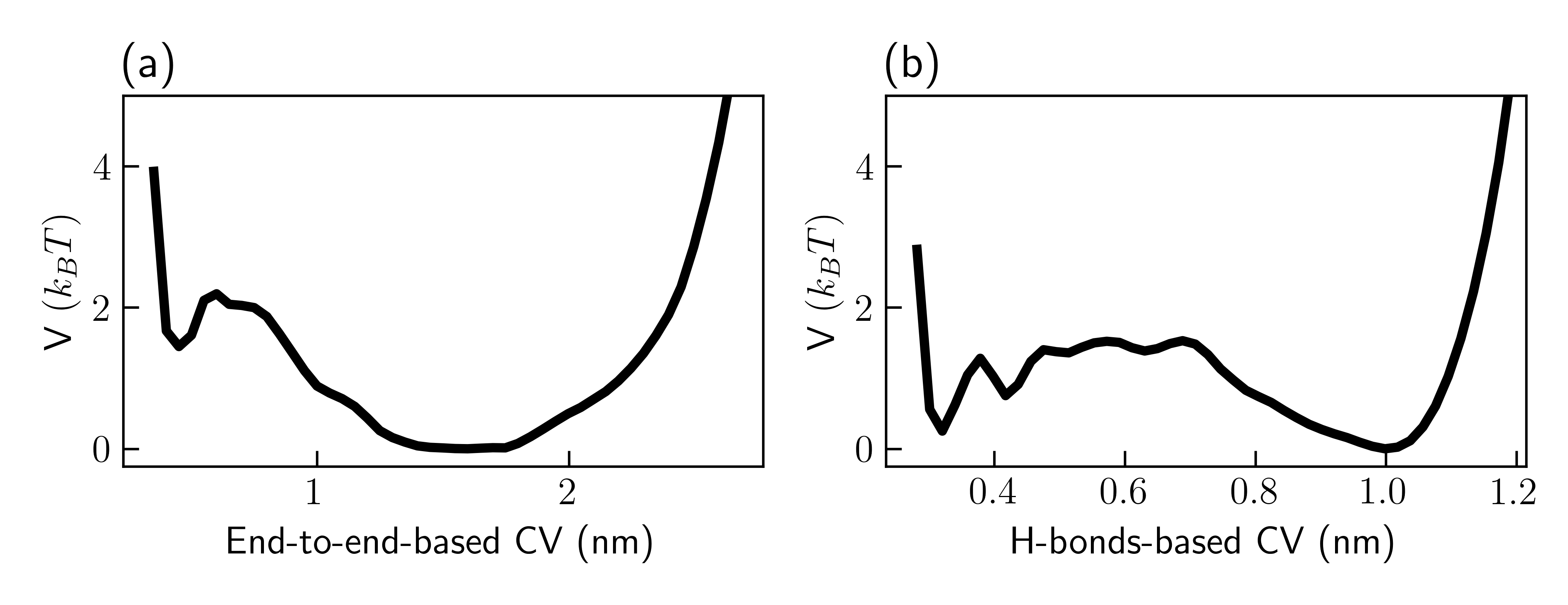}
  \caption{FES of the nine-residues alanine peptide along (a) an end-to-end-based CV and (b) an H-bonds-based CV.}
  \label{fig:1DalanineFES}
\end{figure}

\section{Metadata for Figures 4-5}

Here we report the number of simulations used to produce the bootstrapping results of Figures 4-5 of the main text. 

\begin{table}
\caption{Number of simulations of the three-state potential.}
\label{tab:finiteDifference}
\begin{tabular}{|c|c|c|}
\hline
Timer (ns) & Number of first-passage events & Total number of simulations \\ \hline
0.02 & 163 & 200000 \\ \hline
0.5 & 3453 & 200000 \\ \hline
1 & 5957 & 200000 \\ \hline
2 & 4547 & 100000 \\ \hline
5 & 7366 & 100000 \\ \hline
20 & 17463 & 100000 \\ \hline
\end{tabular}
\end{table}

\begin{table}[!htb]
\caption{Number of simulations of the alanine peptide.}
\label{tab:finiteDifference}
\begin{tabular}{|c|c|c|}
\hline
Timer (ns) & Number of first-passage events & Total number of simulations \\ \hline
1 & 1034 & 10000 \\ \hline
2 & 1621 & 10000 \\ \hline
5 & 2235 & 10000 \\ \hline
10 & 2689 & 10000 \\ \hline
20 & 1621 & 5000 \\ \hline
50 & 2383 & 5000 \\ \hline
\end{tabular}
\end{table}
\bibliography{bibli}